\documentclass[12pt]{article}
\usepackage{url}
\usepackage{graphicx}
\usepackage{float}
\usepackage{sansmath}
\usepackage{xcolor}
\usepackage{setspace}
\usepackage{amsmath}
\usepackage{authblk}
\usepackage{natbib}

\begin{document}
\title{Evaluating probabilistic forecasts of football matches: The case against the Ranked Probability Score}
\author{Edward Wheatcroft}
\affil{London School of Economics and Political Science, Houghton Street, London, United Kingdom, WC2A 2AE.}
\maketitle
\noindent

\begin{abstract}
A scoring rule is a function of a probabilistic forecast and a corresponding outcome that is used to evaluate forecast performance.  A wide range of scoring rules have been defined over time and there is some debate as to which are the most appropriate for evaluating the performance of forecasts of sporting events.  This paper focuses on forecasts of the outcomes of football matches.  The ranked probability score (RPS) is often recommended since it is `sensitive to distance', that is it takes into account the ordering in the outcomes (a home win is `closer' to a draw than it is to an away win, for example).  In this paper, this reasoning is disputed on the basis that it adds nothing in terms of the actual aims of using scoring rules. A related property of scoring rules is locality.  A scoring rule is local if it only takes the probability placed on the outcome into consideration. Two simulation experiments are carried out in the context of football matches to compare the performance of the RPS, which is non-local and sensitive to distance, the Brier score, which is non-local and insensitive to distance, and the ignorance score, which is local and insensitive to distance.  The ignorance score is found to outperform both the RPS and the Brier score, casting doubt on the value of non-locality and sensitivity to distance as properties of scoring rules in this context.
\end{abstract}

\section{Introduction}
Probabilistic forecasting of sporting events such as football matches has become an area of considerable interest in recent years.  One reason for this is that forecasting can help inform gambling decisions and therefore has the potential to support the identification of profitable betting strategies.  Probabilistic forecasting has also grown in popularity in the sports media.  In some media outlets, for example, estimated probabilities are routinely disseminated in match previews and even in-play.  An obvious implication of the growth of probabilistic forecasting in sport is the need for effective methods of forecast evaluation.  This is particularly true in the case of gambling where `beating the bookmaker' is a difficult task which typically requires highly informative predictions.  However, even when the forecasts are used for purposes other than gambling, there is often still an incentive for the forecasts to be informative, or at least to be perceived as such.  There is therefore a need for objective measures of forecast performance.  This paper is concerned with the question of how to evaluate probabilistic forecasts of events such as football matches with three or more possible outcomes. \par

Evaluation of probabilistic forecasts is typically performed using scoring rules, functions of the forecast and corresponding outcome aimed at assessing forecast performance.  A large number of scoring rules have been defined over the years and there is considerable debate surrounding which are the most appropriate.  A common approach with which to differentiate candidate scoring rules is to identify desirable properties and favour scores that have them.  There is often debate, however, surrounding which properties are (most) desirable and hence a lack of consensus remains.  As a result, in fields such as weather forecasting, a wide range of different scores are often presented. \par

One property of scoring rules that is perhaps the most widely agreed upon is called propriety.  A score is proper if, in expectation, it favours a forecast that consists of the distribution from which the outcome is drawn, i.e a perfect probabilistic forecast.  This paper is concerned primarily with two more contentious properties.  One of those properties is locality.  A score is \emph{local} if it only considers the probability at the outcome and disregards the rest of the distribution.  A \emph{non-local} score therefore takes at least some of the rest of the forecast distribution into account. The other property of interest concerns whether a scoring rule takes ordering into account.  Events with discrete outcomes can be divided into two categories: nominal and ordinal.  Ordinal events have a natural ordering.  For example, a question on a survey asking an interviewee to rate a service might have a set of potential responses ranging from `very poor' to `very good'.  It is clear that `very good' ranks higher than `good', whilst `good' ranks higher than `poor'.  Nominal events, on the other hand, have no natural ordering.  For example, there is no obvious way to rank a set of colours or nationalities.  The outcomes of football matches can be considered to be ordinal (along with matches in other sports in which a draw is allowed).  A home win is closer to a draw than it is to an away win.  As such, there is a question of whether a scoring rule should take into account this ordering.  A paper by Constantinou and Fenton argues that forecast probability placed on potential outcomes close to the actual outcome should be rewarded and therefore ordering should be taken into account (\cite{constantinou2012solving}).  Therefore, if the match outcome is a home win, probability placed on a draw should be rewarded more than probability on an away win.  Scoring rules that have this property are referred to as being `sensitive to distance'.  One scoring rule that has this property is the ranked probability score (RPS).  Constantinou and Fenton therefore argue that the RPS is the appropriate score for the evaluation of probabilistic forecasts of football matches. As a result, the RPS has become perhaps the most popular and widely used scoring rule for this purpose.  In this paper, the view that sensitivity to distance in a scoring rule is beneficial is disputed along with Constantinou and Fenton's suggestion that the RPS should be widely used to evaluate football forecasts. \par

Three scoring rules are considered in this paper: the RPS, which is both non-local and sensitive to distance, the Brier score, which is non-local but insensitive to distance and the ignorance score, which is local and therefore also insensitive to distance.  It is argued that the ignorance score is the most appropriate out of these three candidate scores and evidence is presented in the form of two experiments demonstrating that the ignorance score is able to identify a set of perfect forecasts quicker than the other two scoring rules. \par

The question of how probabilistic forecasts of discrete events should be evaluated is one with a long history. An early contribution to the literature was the introduction of the Brier score (\cite{brier1950verification}).  The Brier score considers the squared distance between the forecast probability and the outcome for each possible category in which the outcome could fall (the category in which the outcome falls is represented with a one and all other categories with a zero).  Whilst the Brier score is most commonly applied to binary events, it was originally formulated more generally such that it can be extended to events with more than two possible outcomes.  The ignorance score (\cite{good1992rational,roulston2002evaluating}), often referred to as the logarithmic score, takes a different approach by simply taking the logarithm of the probability placed on the outcome. The rationale behind the ignorance score is in information theory and is closely related to other information measures such as the Kullback-Leibler Divergence (\cite{brocker2007scoring}).  The ranked probability score (\cite{epstein1969scoring}) is closely related to the Brier score but compares the cumulative distribution function of the forecast and the outcome rather than the probability mass function.  Other proposed scoring rules include the spherical score, which combines the probability placed on the outcome with a correction term to ensure that it is proper (\cite{friedman1983effective}) and the quadratic score which simply takes the mean squared distance between the forecast and the outcome (\cite{selten1998axiomatic}).  This paper, however, is concerned only with the ignorance score, Brier score and RPS.  These scoring rules were chosen because we are principally interested in the properties of locality and sensitivity to distance.  The RPS is both non-local and sensitive to distance, the Brier score is non-local and insensitive to distance and the ignorance score is local and insensitive to distance (the ignorance score is in fact the only local and proper scoring rule (\cite{bernardo1979expected}). \par

A range of other properties of scoring rules have been proposed, many of which have been suggested as desirable in some way.  Propriety, as mentioned above is perhaps the most well known property and it stipulates that, in expectation, a scoring rule should favour the distribution from which the outcome was drawn over all others (\cite{brocker2007scoring}).  Another property is locality.  A score is local if only the probability at the outcome is taken into account (\cite{parry2012proper}). Other properties of scoring rules include those that are equitable, defined as those that ascribe the same score, in expectation, to constant forecasts as they do to a random forecast, regular, those that only ascribe an infinite score to a forecast that places zero probability on the outcome (\cite{gneiting2007strictly}), and feasible, those that assign bad scores to forecasts that give material probability to events that are highly unlikely (\cite{maynard}). \par

A number of authors have commented on the value of sensitivity to distance in scoring rules.  \cite{jose2009sensitivity} recommended the use of scoring rules that are sensitive to distance, including for forecasts of football matches.  They provide generalisations of existing scoring rules to make them sensitive to distance.  \cite{stael1970family} also recommended that scoring rules should be sensitive to distance and suggest a family of scoring rules based on the RPS that also have this property.  \cite{murphy1970ranked} compared the formulation of the RPS and the Brier score and recommended that the RPS should at least be used alongside the Brier score when the event of interest is ordered.  \cite{bernardo1979expected}, on the other hand commented that ``when assessing the worthiness of a scientist's final conclusions, only the probability he attaches to a small interval containing the true value should be taken into account.'' arguing for locality as a desirable property. \par 

There is a steadily increasing literature describing methodology for the construction of probabilistic forecasts of sporting events such as football matches (\cite{diniz2019comparing}).  In many of these, scoring rules have been deployed to attempt to assess the quality of those forecasts.  For example, \cite{forrest2005odds} use the Brier score to compare probabilistic forecasts derived from bookmakers' odds and from a statistical model.  \cite{spiegelhalter2009one} use the Brier score to assess the performance of their Premier League match predictions.  The ranked probability score has also been widely used. For example, \cite{koopman2019forecasting} use the RPS to evaluate their dynamic multivariate model of football matches, \cite{baboota2019predictive} use the RPS to evaluate their machine learning approach to football prediction and \cite{schauberger2016modeling} use the RPS alongside cross-validation to select a tuning parameter in their model.  The ignorance score appears to be less widely used than the Brier score and the RPS.  \cite{diniz2019comparing} compare the performance of a number of predictive models using the ignorance score alongside the Brier score and the spherical score whilst it also has been used by \cite{schmidt2008accuracy} alongside the Brier score to assess the probabilistic performance of a prediction market for the 2002 World Cup. \par

This paper is organised as follows.  In section~\ref{section:background}, formal definitions of the scoring rules and their properties are given. In section~\ref{section:rebuttal}, the arguments of Constantinou and Fenton are presented and disputed.  In section~\ref{section:what_scoring_rules_for}, the question of how scoring rules are used in practice is discussed. The philosophical difference between a perfect and imperfect model scenario is discussed in section~\ref{section:perfect_imperfect}.  The performance of the Brier score, Ignorance score and RPS are compared in a model selection experiment using examples from Constantinou and Fenton's paper in section~\ref{section:experiment_one}.  A similar experiment is performed in section~\ref{section:experiment_two} using forecast probabilities derived from bookmakers' odds of actual matches.  Finally, section~\ref{section:discussion} is used for discussion and conclusions. \par

\section{Background} \label{section:background}

\subsection{Definitions of Scoring Rules} 
The three scoring rules considered in this paper are defined as follows. For an event with $r$ possible outcomes, let $p_{j}$ and $o_{j}$ be the forecast probability and outcome at position $j$ where the ordering of the positions is preserved.  The Brier score, generalised for forecasts of events with $r$ possible outcomes, is defined as  
\begin{equation}
\mathrm{Brier}=\sum_{i=1}^{r}(p_{i}-o_{i})^{2}.
\end{equation}
The ranked probability score is defined as
\begin{equation}
\mathrm{RPS}=\sum_{i=1}^{r-1}\sum_{j=1}^{i}(p_{j}-o_{j})^{2}.
\end{equation}
The ignorance score is defined as 
\begin{equation}
\mathrm{IGN}=-\log_{2}(p(Y))
\end{equation}
where $p(Y)$ is the probability placed on the outcome $Y$.

\subsection{Properties of Scoring Rules} \label{section:properties_scoring_rules}
Throughout a long history of research, a large number of properties of scoring rules have been defined.  Here, those that are relevant to the arguments and experiments in this paper are described. \par

Perhaps the most well known property of scoring rules is \emph{propriety}. A score is \emph{proper} if it is optimised, in expectation, with the distribution from which the outcome was drawn.  As such, a proper scoring rule always favours a perfect probabilistic forecast in expectation.  It is widely held that scoring rules that do not have this property should be dismissed (\cite{brocker2007scoring}).  Each of the scoring rules described above are proper. A perfect probabilistic forecast is rarely, if ever, expected to be possible to achieve in practice.  Therefore, in expectation, whilst a proper scoring rule will always rank a perfect forecast more favourably than an imperfect one, different proper scores will often rank pairs of imperfect forecasts differently. \par

Another property of scoring rules is locality.  A score is \emph{local} if it only takes into account the probability at the outcome.  If any of the rest of the distribution is taken into account by the score, it is non-local. The ignorance score is local whilst the Brier Score and RPS are both non-local. \par

For discrete events, another property concerns whether the score takes into account the ordering of a set of potential outcomes.  Scores that do this are defined as \emph{sensitive to distance}.  For sporting events, for example, a draw and a home win can be considered to be closer together than a home win and an away win.  An scoring rule that is sensitive to distance will therefore reward probability placed on an event closer to the actual outcome.  Whilst the RPS is sensitive to distance, the ignorance and Brier Scores are not since they do not take into account the ordering of the possible outcomes. \par

\section{A Rebuttal of Arguments in Favour of the RPS} \label{section:rebuttal}
The popularity of the RPS for evaluating probabilistic forecasts of football matches is largely due to a paper written by Constantinou and Fenton, published in the Journal of Quantitative Analysis in Sports in 2012 (\cite{constantinou2012solving}).  The crux of the argument in that paper is that probability placed on potential outcomes `close' to the actual outcome should be rewarded more than probability placed on those that are `further away'.  If the home team is currently winning by one goal, it would take the away team to score one more goal for the match to end in a draw and two more goals for it to end in an away win and, therefore, the potential outcomes are, in a sense, ordered. The authors claim that, in light of this, only scoring rules that are sensitive to distance should be considered.  A natural choice is therefore argued to be the RPS. \par

With the aim of presenting further evidence towards the suitability of the RPS, Constantinou and Fenton define five hypothetical football matches, each with a specified outcome (i.e. a home win, a draw or an away win).  For each match, they define a competing pair of probabilistic forecasts and use general reasoning to argue that, given the defined outcome, one is more informative than the other.  They then show that the RPS is the only scoring rule out of a number of candidates that assigns the best score to their favoured forecast in each case, and argue that this provides evidence of its suitability.  We dispute the validity of this reasoning.  We argue that the approach by which the performance of the scores is compared under a specific outcome of the match is flawed. Instead, scores should be compared by considering the underlying probability of each possible outcome, thereby taking into account the underlying probability distribution of the match. This is a much more difficult task. \par

To provide a setting with which to illustrate the arguments of Constantinou and Fenton and to provide counterarguments, details of the five hypothetical matches used as examples in that paper are reproduced.  The outcome of each match, the two forecasts and an indicator of Constantinou and Fenton's favoured forecast ($\alpha$ or $\beta$) are shown in table~\ref{table:constantinou_scenarios}.

\begin{table}[]
\begin{tabular}{|l|l|lll|l|l|}
\hline
Match & Forecast    & p(H) & p(D) & p(A) & Result & `Best' Forecast \\
\hline
1     & $\alpha$ & 1    & 0    & 0    & H      & $\alpha$   \\
      & $\beta$  & 0.9  & 0.1  & 0    &        &            \\
\hline
2     & $\alpha$ & 0.8  & 0.1  & 0.1  & H      & $\alpha$   \\
      & $\beta$  & 0.5  & 0.25 & 0.25 &        &            \\
\hline
3     & $\alpha$ & 0.35 & 0.3  & 0.35 & D      & $\alpha$   \\
      & $\beta$  & 0.6  & 0.3  & 0.1  &        &            \\
\hline
4     & $\alpha$ & 0.6  & 0.25 & 0.15 & H      & $\alpha$   \\
      & $\beta$  & 0.6  & 0.15 & 0.25 &        &            \\
\hline
5     & $\alpha$ & 0.57 & 0.33 & 0.1  & H      & $\alpha$   \\
      & $\beta$  & 0.6  & 0.2  & 0.2  &        &            \\
\hline
\end{tabular}
\label{table:constantinou_scenarios}
\caption{Match examples defined by Constantinou and Fenton.  In each case, forecast $\alpha$ is described as superior to forecast $\beta$ by the authors.}
\end{table}

The reasoning given by the authors for favouring each forecast is as follows.  For match one, forecast $\alpha$ predicts a home win with total certainty and must outperform any other forecast (including $\beta$).  For match two, forecast $\alpha$ places more probability on the outcome than forecast $\beta$ and therefore provides the most informative forecast.  For match three, the only match in which the outcome is defined to be a draw, it is argued that, although both forecasts place the same probability on the outcome, forecast $\alpha$ should be favoured because the probability placed on a home win and an away win are evenly distributed and therefore `more indicative of a draw'.  For match four, it is argued that forecast $\alpha$ should be favoured because, although both forecasts place the same probability on the outcome, $\alpha$ places more probability `close' to the home win, i.e. on a draw, than $\beta$ and therefore is more favourable.  Finally, for match five, which is described as the most contentious case, whilst $\beta$ places more probability on the outcome than $\alpha$, the authors argue that forecast $\alpha$ is, in fact, more desirable than $\beta$ because it is `more indicative of a home win' due to the greater probability placed on the draw.  To provide further justification for favouring forecast $\alpha$, they give an example in which a gambler uses the forecast to inform a bet on the binary event of whether the match ends with any outcome other than an away win (commonly known as a lay bet).  Since the sum of the probabilities on the home win and the draw are higher for forecast $\alpha$ than for forecast $\beta$, they suggest that $\alpha$ is more desirable for that purpose. \par

Our main counterargument to the reasoning of Constantinou and Fenton concerns their assertion that forecast $\alpha$ outperforms forecast $\beta$ in each case.  In fact, it is impossible to say which of the two forecasts should be preferred in each case without considering the underlying probability distribution of the match (which is unknown in practice).  Consider match one.  Here,  they argue that forecast $\alpha$ should be rewarded more than any other forecast since it predicts the outcome with absolute certainty.  This seems entirely reasonable since no forecast is able to place more probability on the outcome.  However, it does \emph{not} follow from this that $\alpha$ is the best forecast.  To illustrate this, consider the case in which $\beta$ represents the true underlying probability distribution of the match; i.e. the match will end with a home win with probability $0.9$, a draw with probability $0.1$ and an away win with probability $0$.  It is not contentious to state that $\beta$ is the best forecast in this setting and we argue that it would be deeply flawed to claim otherwise.  Forecast $\alpha$ should not be considered to be the best forecast simply because the match happened to end in a home win (which would happen with 90 percent probability in this case).  In a succession of football matches in which the underlying probability is represented by $\beta$ and the forecast is $\alpha$, a draw would eventually occur, with forecast $\alpha$ placing zero probability on that event.  The same logic can be applied if the underlying probability distribution is represented by forecast $\alpha$, in which case, $\alpha$ can objectively be considered to be the best forecast.  In summary, without knowing the underlying probability distribution of the match, the answer to the question of which forecast is best can only be `it depends'.  In practice, of course, it is never possible to know the underlying distribution and therefore we cannot distinguish the performance of the two forecasts on the basis of a single match. \par

\begin{figure}[!htb]
    \centering
    \includegraphics[scale=0.35]{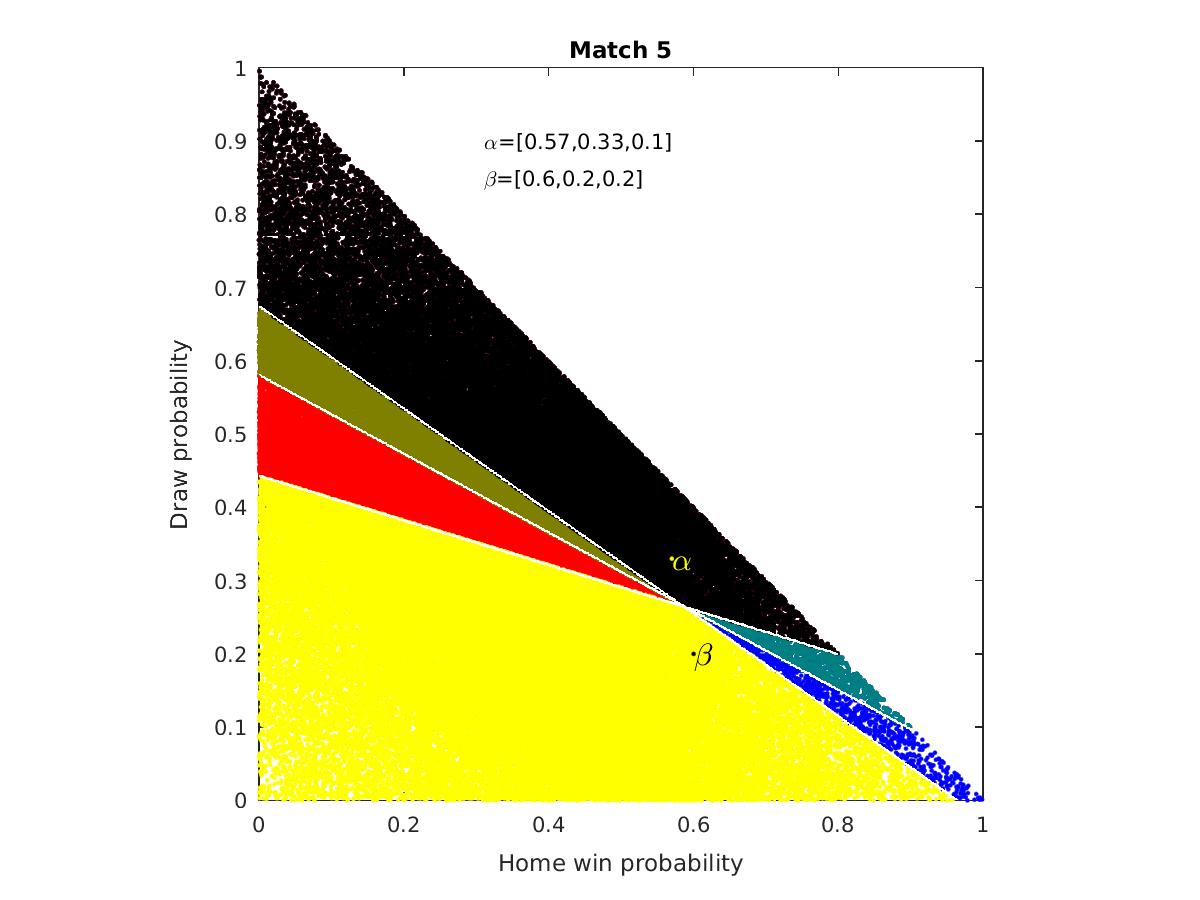}
    \caption{Randomly chosen probability distributions of match five coloured according to which forecast ($\alpha$ or $\beta$) is preferred by each of the three scoring rules.  The colour scheme is described in table~\ref{table:key_for_alpha_beta_plot}.}
    \label{figure:match5_imperfect_scores}
\end{figure}

The effect of the probability distribution of the match on the favoured forecast under each score is now demonstrated.  For a given probability distribution, the expected score of each of forecasts $\alpha$ and $\beta$ are calculated, in order to determine which is preferred by each scoring rule.  This is repeated for a large number of randomly selected underlying probability distributions. This is demonstrated for match five in figure~\ref{figure:match5_imperfect_scores}.  Here, each dot represents a different probability distribution of the match with the probability of a home win and a draw on the $x$ and $y$ axes respectively.  Each dot is coloured according to which of the two candidate forecasts is preferred under the three scoring rules. The colour scheme is defined in table~\ref{table:key_for_alpha_beta_plot}.  For example, if a point is coloured blue, the RPS prefers $\alpha$ whilst the ignorance and Brier scores prefer $\beta$ under that distribution.  Whilst the colour scheme might seem difficult to interpret at first, it becomes much clearer when it is considered that points coloured green, blue and red represent distributions in which only the ignorance, RPS and Brier score prefer $\alpha$ and that the colours of overlapping regions are defined by mixing those colours. Note that, for this particular match, there are no green areas, that is there are no underlying distributions in which the ignorance score prefers $\alpha$ and the RPS and Brier score prefer $\beta$. \par

\begin{table}[]
\begin{tabular}{|l|c|c|c|}
\hline
Colour & Ignorance & RPS & Brier \\
\hline
Green & $\alpha$ & $\beta$ & $\beta$ \\
Blue & $\beta$ & $\alpha$ & $\beta$ \\
Red & $\beta$ & $\beta$ & $\alpha$ \\
Turquoise & $\alpha$ & $\alpha$ & $\beta$ \\
Brown & $\alpha$ & $\beta$ & $\alpha$ \\
Purple & $\beta$ & $\alpha$ & $\alpha$ \\
Yellow & $\beta$ & $\beta$ & $\beta$ \\
Black & $\alpha$ & $\alpha$ & $\alpha$ \\
\hline
\end{tabular}
\label{table:key_for_alpha_beta_plot}
\caption{Colour scheme for figures~\ref{figure:match5_imperfect_scores} and~\ref{figure:match_1_to_4_imperfect_scores}.}
\end{table}

The first conclusion to be drawn from figure~\ref{figure:match5_imperfect_scores} is that, clearly, as previously discussed, the forecast favoured by each scoring rule depends on the underlying probability distribution. Moreover, the choice of scoring rule impacts which of the two forecasts is preferred.  We can look at each of the regions and try to understand how and why the three scoring rules differ.  Consider the blue region in the bottom right of the figure.  A point located in the very bottom right represents a probability distribution which places a probability of one on a home win and therefore zero on both a draw and an away win.  Here, the RPS is the only score that favours $\alpha$ over $\beta$.  This seems somewhat counterintuitive and can be argued to be a weakness of the scoring rule.  Here, the RPS rewards probability placed on a draw, regardless of the fact that that outcome \emph{cannot} happen.  The cost of doing this is that, out of the two forecasts, the one that places less probability on the outcome is favoured. \par

\begin{figure}[!htb]
    \centering
    \includegraphics[scale=0.35]{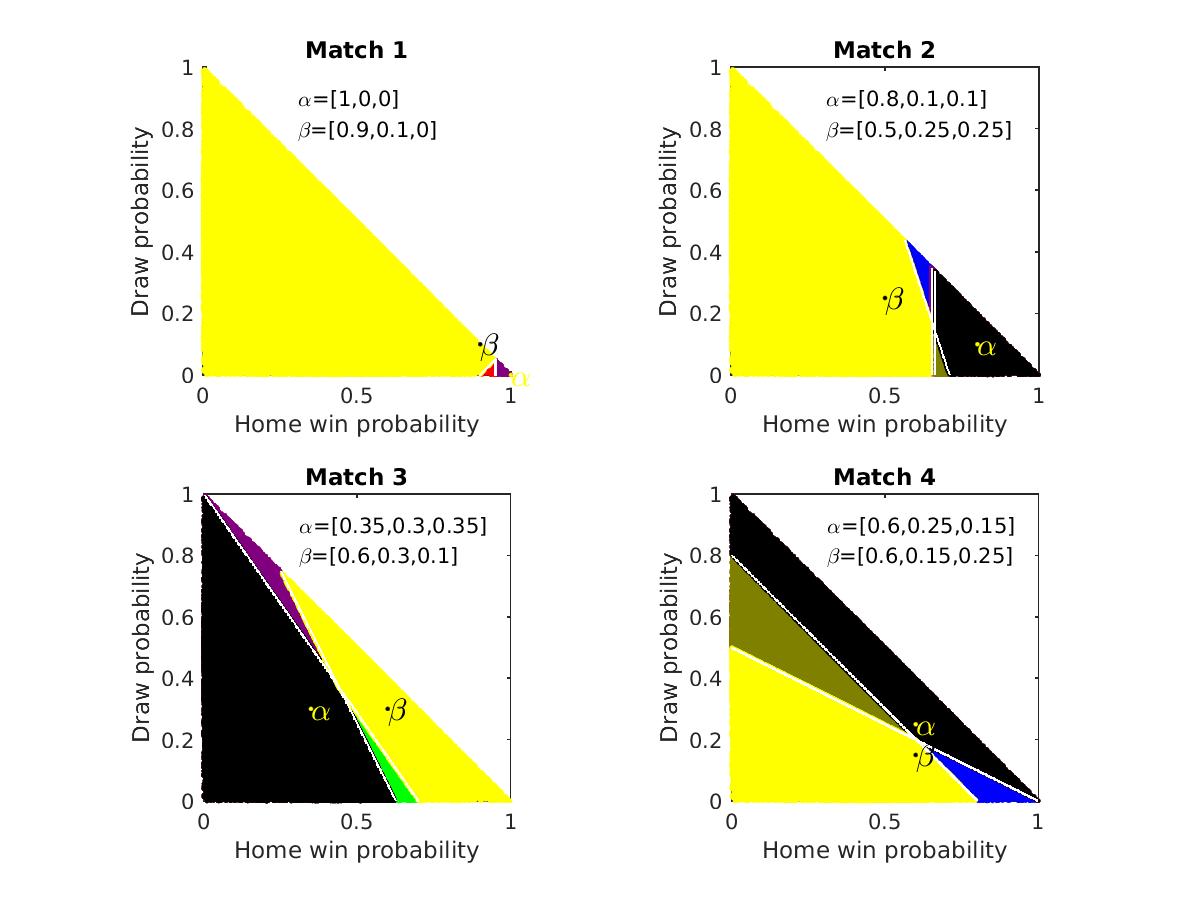}
    \caption{Randomly chosen probability distributions of matches one to four coloured according to which forecast ($\alpha$ or $\beta$) is preferred by each of the three scoring rules.  The colour scheme is described in table~\ref{table:key_for_alpha_beta_plot}.}
    \label{figure:match_1_to_4_imperfect_scores}
\end{figure}

The same information as shown in figure~\ref{figure:match5_imperfect_scores} for match five is shown for matches one to four in figure~\ref{figure:match_1_to_4_imperfect_scores}.  This reinforces the importance of the underlying probability distribution and how the choice of forecast depends heavily on the scoring rule. \par

In practice, scoring rules are usually used to assess the performance of forecasting \emph{systems} rather than individual forecasts.  A forecasting system is a set of rules that is used to generate forecasts of different events in some common way.  For example, a forecasting system might be built on the basis of an individual model, a combination of models or the judgement of a particular person and can be applied to generate forecasts of a range of events (e.g. football matches).  Forecasting systems are then evaluated by taking the average score over many events according to some scoring rule.  This provides a basis with which to select a forecasting system for the prediction of future events. \par

Before moving on, it is of interest to address two particular points made by Constantinou and Fenton in favour of forecast $\alpha$ for match five.  Firstly, they describe a situation in which the forecasts are used to inform a `lay' bet on an away win.  They argue that, since the combined probability placed on a home win or a draw is higher for forecast $\alpha$ than for forecast $\beta$, $\alpha$ is a better forecast, given this outcome.  There is a simple counterargument to this.  If a gambler intends to use the forecasts to make lay bets such as the one described, the resulting binary forecasts formed by adding the home win and draw probabilities should be evaluated separately.  This is because the new binary forecasts take a different form and have a different aim.  It does not makes sense during evaluation to attempt to pre-empt how the forecasts might be used to create other forecasts of a different nature.  In fact, the original match outcome forecasts and the binary forecasts might even favour a different forecasting system. For example, one forecasting system might be poor at distinguishing a home win from a draw but good at estimating the probability of an away win.  Tying one's hands to create and use a one size fits all forecast seems unnecessary and counterproductive in this case. \par

The second point of contention regards the `indicativeness of a home win' in match five.  The authors argue that despite the fact that forecast $\beta$ places more probability on the outcome than forecast $\alpha$, forecast $\alpha$ is more indicative of a home win, due to the increased probability placed on the draw.  It should be noted here that, were the probability on the draw reduced to $0.3$ and the probability on the away win increased to $0.23$, the RPS would favour forecast $\beta$ and thus the `indicativeness' of a home win is somewhat arbitrary. \par 

The primary claim of Constantinou and Fenton is that probability placed on possible outcomes that are `close' to the actual outcome should be rewarded more than probability placed on outcomes that are `further away'.  Furthermore, they argue that the RPS provides a scoring rule that does this and is therefore suitable for evaluating forecasts of football matches.  However, as described above, by not considering the underlying distribution of the match, it is not possible to state that one forecast is better than another and therefore the reasoning given in support of the RPS does not provide a compelling argument.  We therefore consider the question of which forecast is assigned the best score, when conditioned on a single outcome, to be moot and we do not consider it further.  Instead, we define potential goals of using scoring rules and ask whether the sensitivity to distance property offered by the RPS has any value in achieving them. \par

\section{What are Scoring Rules For?} \label{section:what_scoring_rules_for}
The principle intention of this paper is to assess the value of scoring rules that are non-local and sensitive to distance in the context of forecasts of football matches. In order to attempt to assess the merits of these properties, it is useful to consider the aims behind the deployment of scoring rules.  For the properties of interest to have value, there should be some practical benefit in terms of achieving those aims.  Here, we discuss the aims behind the application of scoring rules with a view to assessing  whether the non-local and sensitivity to distance properties help to achieve them. \par

One obvious aim of scoring rules is to provide a means of comparison between competing forecasting systems.  There are many contexts in which one might want to make such comparisons.  One might have a finite set of competing probabilistic forecasts of the same events and be looking to determine which is the most informative.  For example, a broadcaster may want to decide which forecasts are most useful to show in its sports coverage or a gambler may wish to decide which forecasting service to subscribe to in order to aid their betting decisions. A means of comparison can also be important in the context of model development.  A forecaster looking to improve the performance of their forecasting system by, for example, increasing the number of factors included in the model, may want to assess whether these changes result in improved forecasts.  Parameter selection also falls under the umbrella of forecast comparison since each set of parameter values will lead to a different set of forecasts. Since parameters usually take continuous values, parameter selection can be considered to be a comparison between an infinite number of sets of forecasts. \par

Whilst selecting one of two or more sets of forecasts may be considered to be important in a range of settings, this alone does not give an indication of the magnitude of the difference in skill.  An additional question of interest concerns how much more informative one set of forecasts is over another and whether this difference is significant.  Typically, two sets of forecasts are compared using the difference in their mean score (\cite{wheatcroft2019interpreting}). Resampling techniques can then be used to determine if that difference is significant.  An interesting question concerns whether the difference in scores has an interpretation in terms of the relative performance of the forecasting systems of interest.  In fact, to our knowledge, only one of the scoring rules considered in this paper has a useful interpretation when considered in this way and that is the ignorance score.  The difference between the mean ignorance scores of two forecasting systems represents the difference in information provided by each one expressed in bits.  This means that calculating $2$ to the power of the mean relative ignorance between forecasting systems one and two yields the mean increase in probability density placed on the outcome by the former over the latter. \par
 
Whilst scoring rules are useful tools for evaluating and comparing forecasting systems, it is important to acknowledge their limitations.  Often, a set of forecasts are used with a specific purpose in mind.  In sports forecasting, this might be to aid a decision whether to place a bet on a certain outcome, whether to select a certain player for a match or which play to make during a game.  It is therefore crucial to determine whether the forecasts are fit for that specific purpose.  For example, a set of forecasts may successfully incorporate important information and therefore score better than alternative forecasts that do not incorporate this information yet still not be fit for a specific purpose.  For example, using a set of forecasts to choose whether to place bets may result in a substantial loss which would have been avoided had the bets not been placed at all. \par

\section{Perfect and imperfect model scenarios} \label{section:perfect_imperfect}
Philosophical approaches to the comparison of scoring rules typically consider two distinct settings: the perfect model scenario, in which one of the candidate forecasting systems coincides with the probability distribution that generated the outcome (often referred to as the data generating model (DGM)) and the imperfect model scenario, in which each candidate forecasting system is imperfect (\cite{judd2001indistinguishable_perfect,judd2004indistinguishable}). In the perfect model scenario, there should be no ambiguity as to which set of forecasts is most desirable; a perfect forecasting system is always better than an imperfect one.  In the latter, on the other hand, the `best' forecasting system is subjective and the question of which one is the most desirable is also subjective.  \par

In the perfect model setting, there are two directly linked questions of interest.  Firstly, `does the scoring rule always favour the perfect forecasting system in expectation?'  Scoring rules that do this are called proper and it is generally considered that a chosen scoring rule should have this property (\cite{brocker2007scoring}).  As discussed in section~\ref{section:properties_scoring_rules}, each of the three scoring rules considered in this paper are proper and thus they cannot be distinguished in this way. A closely linked means of comparison for scoring rules assumes that each one is proper and assesses how many past forecasts and outcomes are required to have a given probability of selecting the perfect forecasting system. Requiring fewer forecasts and outcomes to do this means that the information is used more efficiently and therefore that there is a better chance of selecting the best forecasting system for future events.  This observation forms the basis of the experiments presented in this paper. \par

In practice, one can never expect any of the candidate forecasting systems to be perfect and therefore the perfect model case is generally accepted to be only a theoretical construct. It can nonetheless be argued that the performance of scoring rules in this context is important. If, in expectation, a scoring rule does not favour a perfect forecasting system over all others, one should be uneasy about the ability of that scoring rule to favour useful imperfect forecasting systems over misleading ones.  Similarly, the efficiency in which a scoring rule uses the information in past forecasts and outcomes ought to tell us something about the way in which each scoring rule uses the information provided to it.  For example, in the context of non-local scoring rules that are sensitive to distance, if these properties are truely useful, we might expect that that extra information should be capable of distinguishing perfect and imperfect forecasting systems more quickly. \par

In practical situations, since none of the candidate forecasting systems are expected to be perfect, all exercises in forecasting system selection fall into the imperfect category. In this setting, unlike the perfect model case, proper scores will often favour different imperfect forecasting systems. Distinguishing the scoring rules is then a question of identifying which type of imperfect forecasts should be preferred.  Other than the analysis demonstrated in figures~\ref{figure:match5_imperfect_scores} and~\ref{figure:match_1_to_4_imperfect_scores}, this question is left as future work.  \par

\section{Experiment one - Repeated outcomes of the same match} \label{section:experiment_one}
In this experiment, the five pairs of forecasts defined by Constantinou and Fenton and shown in table~\ref{table:constantinou_scenarios} are used to assess the probability that each of the three candidate scoring rules identifies a perfect forecasting system over an imperfect one for a given number of past forecasts and outcomes. For a given pair of forecasts, define the outcomes of a series of $n$ matches by drawing from forecast $\alpha$ or forecast $\beta$ with equal probability $0.5$.  Define two forecasting systems as follows.  The \emph{perfect forecasting system} always knows which of the two distributions from which the outcome is drawn and therefore always defines the correct distribution as the forecast.  The \emph{imperfect forecasting system}, on the other hand, always issues the alternative distribution as the forecast.  A scoring rule is defined to `select' a forecasting system if it is assigned the lowest mean score over $n$ forecasts.  The probability that each scoring will select the perfect forecasting system is calculated for different values of $n$ and the experiment is carried out using each of the forecast pairs in table~\ref{table:constantinou_scenarios}. 

\subsection{Results}
\begin{figure}[!htb]
    \centering
    \includegraphics[scale=0.6]{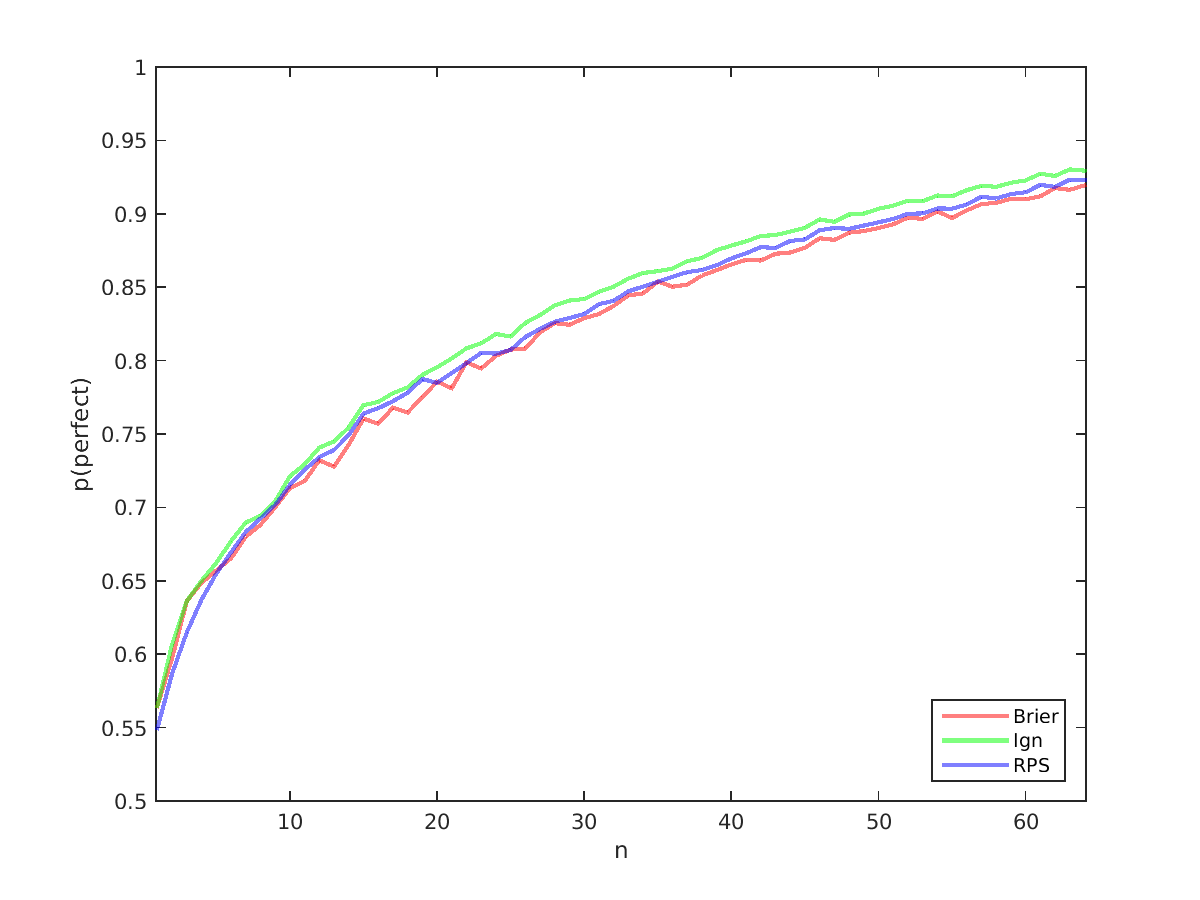}
    \caption{Probability of each scoring rule selecting the perfect forecasting system as a function of $n$ for match 5.}
    \label{figure:prob_perfect_fun_n_match_5}
\end{figure}

Perhaps the most interesting of the five examples defined by Constantinou and Fenton is match five.  Here, both forecast $\alpha$ and forecast $\beta$ place similar probability on a home win but forecast $\alpha$ places more probability on the draw.  The probability of each scoring rule identifying the perfect forecasting system in this case is shown as a function of $n$ in figure~\ref{figure:prob_perfect_fun_n_match_5}.  Here, the ignorance score outperforms both the Brier score and the RPS for almost every tested value of $n$, whilst there is little difference in the performance of the Brier score and RPS. \par

The results for matches one to four are shown in figure~\ref{figure:prob_perfect_fun_n_match_5}. For match one, the ignorance score clearly outperforms both the Brier score and the RPS for relatively large values of $n$, whilst the difference is minimal for lower values. The non monotonic nature of the probabilities under the RPS and Brier scores may seem surprising at first but, in fact, can easily be explained.  All three scoring rules punish the imperfect forecasting system when the outcome is a draw, since the forecast in this case predicts a home win with certainty.  The overall probability of a draw for a given realisation is $0.05$ since the probability that the outcome is drawn from $\beta$ is $0.5$ and the probability of a draw given $\beta$ is $0.1$.  For all values of $n$ less than $20$, once a draw has occurred, no combinations of other outcomes can result in the imperfect forecasting system being assigned a better mean score than the perfect forecasting system.  When $n$ is greater than $20$, on the other hand, the imperfect forecasting system can still `recover' from such a situation as long as there is only one such occurrence.  The probability for the ignorance score, on the other hand, is monotonic and this is because the ignorance score assigns an infinitely bad score to a forecast that places zero probability on an outcome.  Therefore, once such a case has been observed, the imperfect forecasting system cannot achieve a better score than the perfect forecasting system and, since the probability of observing such a case increases with $n$, the probability is monotonically increasing. \par

For match two, whilst the probabilities of selecting the perfect forecasting system are similar for each score, the ignorance score slightly outperforms the other two scores for all values of $n$.  In terms of the Brier score and the RPS, neither appears to be systematically better than the other. \par

For match three, the ignorance score tends to outperform the other two scores for all $n$ greater than three, whilst, again, there is no obvious systematic difference between the performance of the RPS and Brier scores.  For very small $n$, the ignorance score achieves a lower probability of selecting the perfect forecasting system. However, caution should be applied in such cases since scoring rules are designed with a relatively large number of forecast pairs in mind, that is, if the aim were to apply them to small $n$, they might be designed differently. \par

Match four provides perhaps the most interesting results.  In this case, $\alpha$ and $\beta$ differ only in the probabilities placed on a draw and an away win.  There is therefore only a small difference between the perfect and imperfect forecasting systems.  This is reflected in the fact that the probability of choosing the perfect forecasting system increases relatively slowly with $n$.  Here, whilst the ignorance and Brier scores perform similarly well, there is a distinct advantage for both over the RPS.  Whilst the RPS performs relatively well for low $n$, the value of increasing $n$ is far lower than for the ignorance and Brier scores, i.e. the RPS does not make good use of the extra information provided by increasing the number of forecasts and outcomes. \par

Overall, from these results, there is no evidence that the RPS outperforms either the Brier or ignorance scores and, in fact, there is some evidence that the opposite is true.  The RPS does not typically make good use of additional sample members in comparison to the other two scores.  Looking more closely at the results, the stark difference in performance in match four, and to some extent match five, suggests that the biggest difference in performance might be in cases in which the difference between the two candidate forecasts is relatively small and this observation provides a motivation for the design of experiment two.  Experiment one considers only a case with repeated forecasts from one of two candidate distributions.  In practice, there is usually interest in forecasts of different events rather than a large number of realisations of the same event.  In experiment two, the performance of the three candidate scoring rules is compared in the context of forecasts of a wide range of different football matches generated from actual bookmakers' odds. \par

\begin{figure}[!htb]
    \centering
    \includegraphics[scale=0.6]{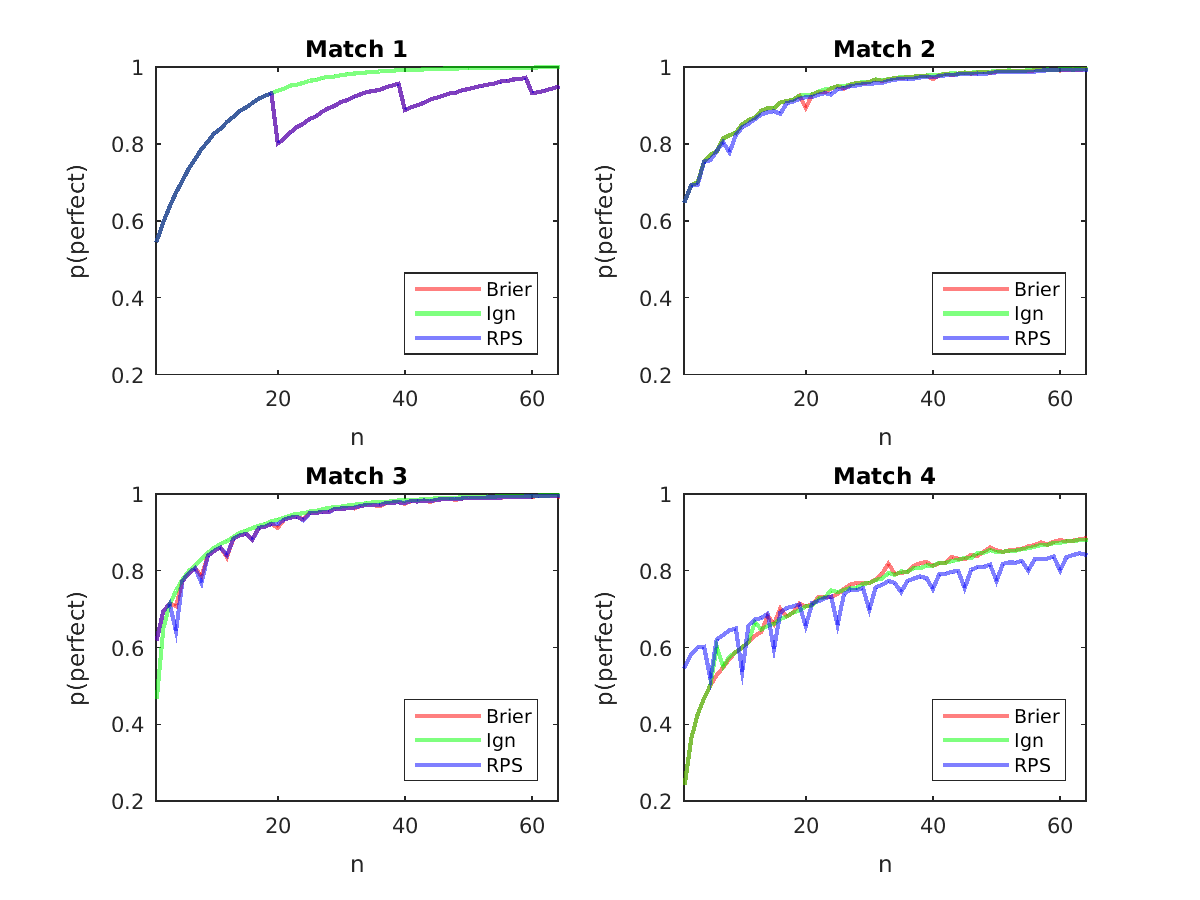}
    \caption{Probability of each scoring rule selecting the perfect forecasting system as a function of $n$ for matches 1 to 4.}
    \label{figure:prob_perfect_fun_n_match_1_to_4}
\end{figure}

\section{Experiment two - forecasts based on match odds} \label{section:experiment_two}
In experiment two, the aim is to assess the effectiveness of each scoring rule in terms of distinguishing a `perfect' forecasting system from an `imperfect' one in a more realistic setting in which each match has a different probability distribution.  To do this, artificial pairs of forecasts are created in which one represents the true distribution and the other is imperfect.  The aim is then to estimate the probability that each scoring rule selects the set of true distributions. \par

To obtain sets of forecasts that are realistic in terms of actual football matches, bookmakers' odds on past matches are used which are converted into probabilistic forecasts.  These are taken from the repository of football data at \url{football-data.co.uk} which supplies free-to-access data from a range of European leagues.  Details of the data and how the odds are used to generate probabilistic forecasts are given in the appendix.  Odds from a total of 39,343 matches are available and form the basis of a set of candidate probability distributions. \par

We seek $n$ pairs of forecasts such that one represents the true distribution of the outcome, and therefore a perfect forecast, whilst the other represents an imperfect forecast.  In order to test the effect of different levels of imperfection, we define a method of controlling it.  To create a perfect forecast and corresponding outcome, a distribution is randomly drawn from the candidate set and defined to be the perfect forecast.  A random draw from that distribution is then taken and defined to be the outcome. Next, we seek an alternative, imperfect forecast from the candidate set.  Here, we apply a condition on the similarity of the candidate forecasts with the perfect forecast. Let the perfect forecast be defined by $\{p_{h},p_{d},p_{a}\}$ where $p_{h}$, $p_{d}$ and $p_{a}$ represents the forecast probability of a home win, draw and away win respectively.  For each forecast in the candidate set, define the `distance' from the true probability distribution to be
\begin{equation}
\epsilon=\frac{1}{3}(|\tilde{p}_{h}-p_{h}|+|\tilde{p}_{d}-p_{d}|+|\tilde{p}_{a}-p_{a}|).
\end{equation}
We define some threshold value $\delta$, find all forecasts for which $\epsilon$ is less than $\delta$ (excluding the perfect forecast itself) and randomly draw the imperfect forecast from that set. This process is repeated $n$ times such that there are a total of $n$ pairs of forecasts. We define the `perfect forecasting system' to be the system that always issues the perfect forecast from the pair and the `imperfect forecasting system' to be such that the alternative, imperfect forecast is always issued.  The experiment is repeated for multiple values of $n$ and different levels of the parameter $\delta$, which governs the imperfection. \par

\subsection{Results}
\begin{figure}[!htb]
    \centering
    \includegraphics[scale=0.6]{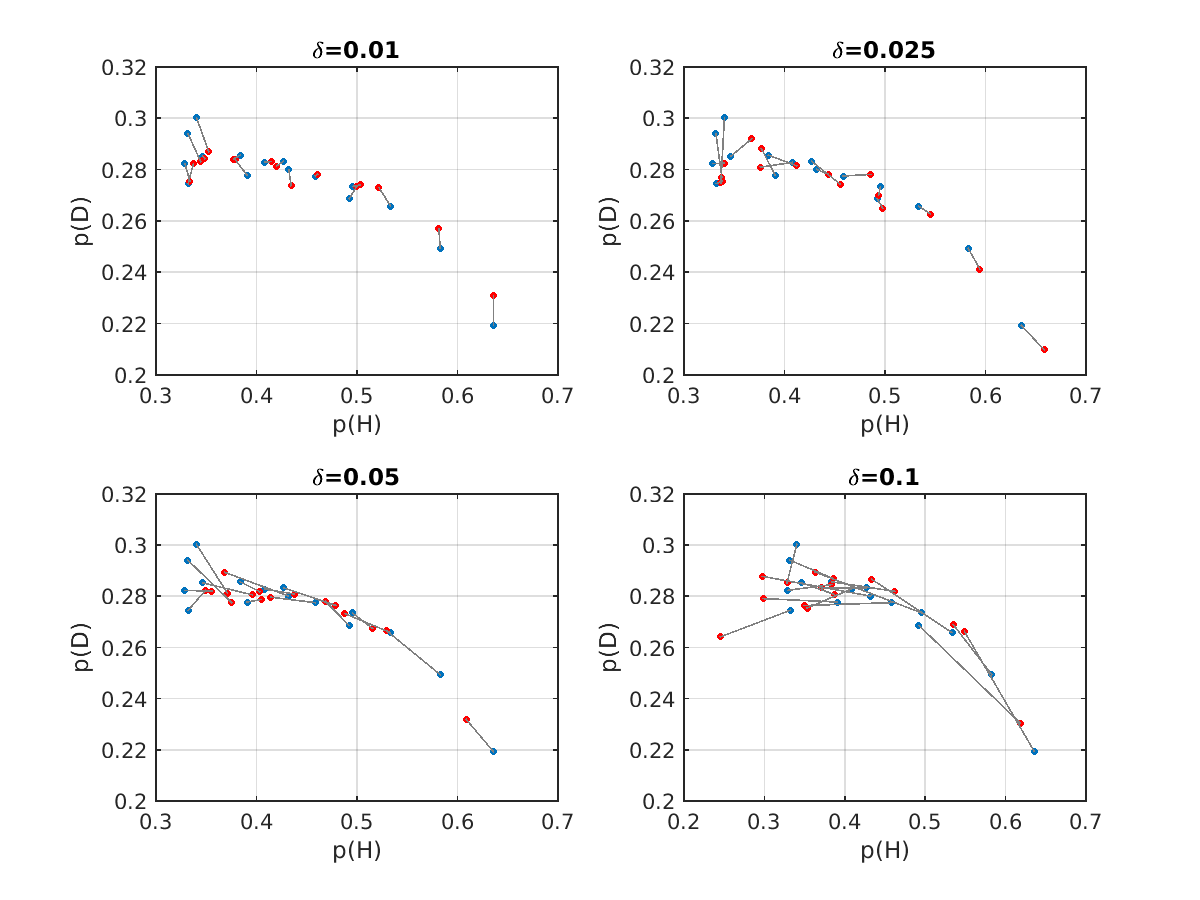}
    \caption{Examples of forecast pairs for different levels of imperfection.  The probability placed on a home win and a draw is represented by the $x$ and $y$ axes respectively.  The grey lines join pairs of forecasts for the purpose of the experiment.}
    \label{figure:model_error_plot}
\end{figure}

The effect of different levels of imperfection, governed by the selected value of $\delta$, is demonstrated in figure~\ref{figure:model_error_plot}.   Each blue dot represents a perfect forecast, with the $x$ and $y$ axes representing the probability of a home win and a draw respectively. The grey line links each of these with the corresponding imperfect forecast.  Increasing the value of $\delta$ tends to result in more distinct pairs of forecasts and therefore a higher level of imperfection. \par

\begin{figure}[!htb]
    \centering
    \includegraphics[scale=0.6]{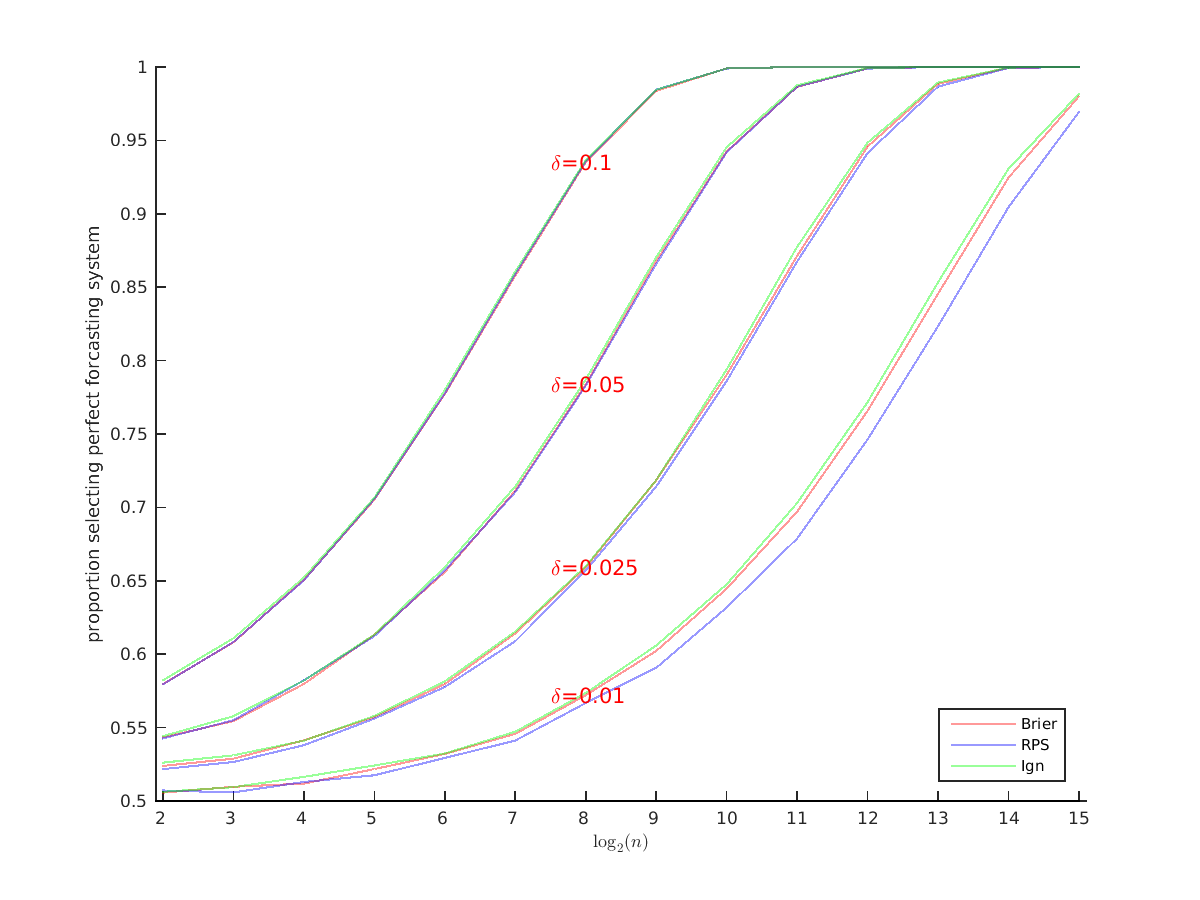}
    \caption{The proportion of cases in which the perfect forecasting system is selected by each scoring rule system as a function of $\log_{2}(n)$ for different values of $\delta$.}
    \label{figure:prop_correct}
\end{figure}

\begin{figure}[!htb]
    \centering
    \includegraphics[scale=0.6]{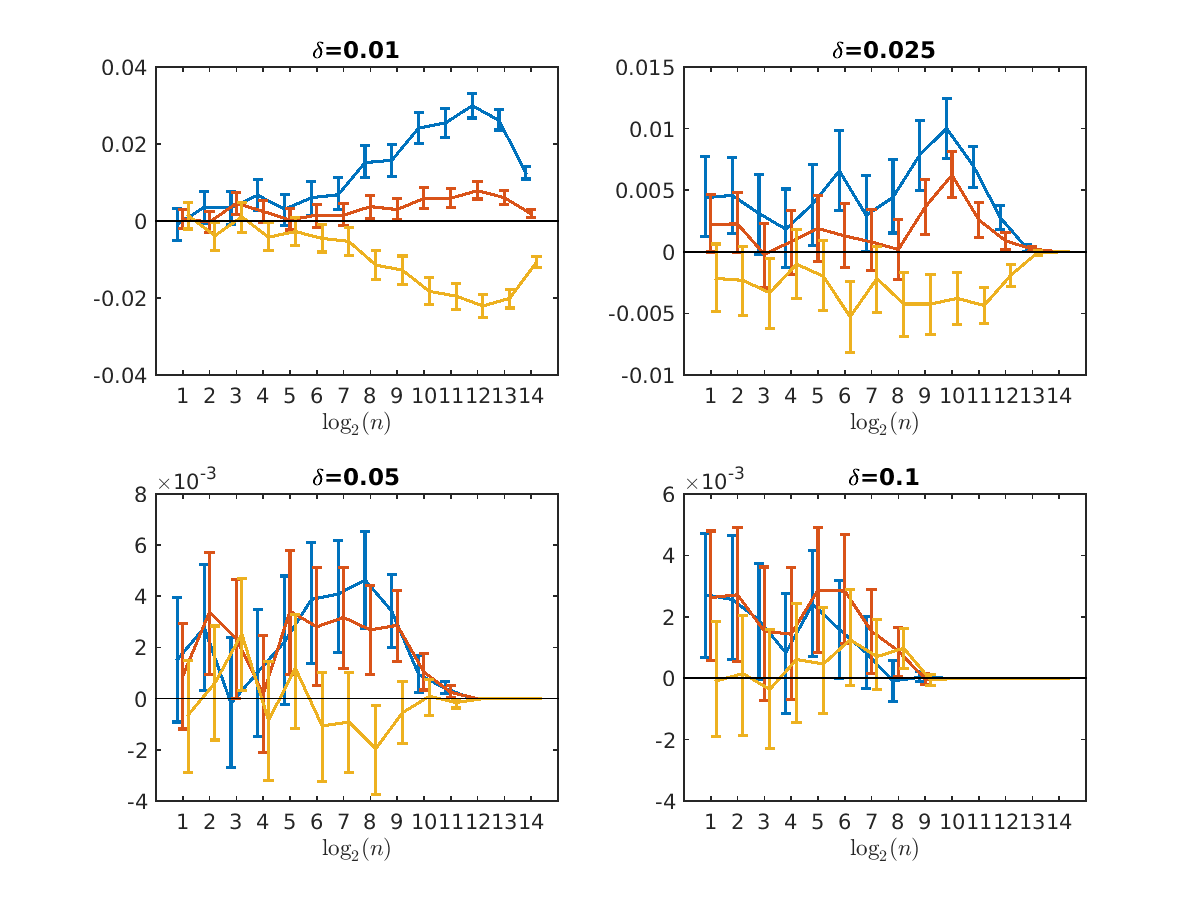}
    \caption{Pairwise differences in the proportion of cases in which the perfect forecasting system is selected between the ignorance and RPS (blue), ignorance and Brier score (red) and Brier score and RPS (yellow) as a function of $\log_{2}(n)$ with 95 percent resampling intervals of the mean.}
    \label{figure:Diff_wins_error_bars}
\end{figure}

The proportion of forecast pairs in which the perfect forecasting system is selected over the imperfect forecasting system is shown for each score and value of $\delta$ as a function of $\log_{2}(n)$ in figure~\ref{figure:prop_correct}.  The red, blue and green lines represent this proportion for the Brier score, Ignorance score and RPS respectively for the stated value of $\delta$.  For higher levels of imperfection, that is when $\delta$ is high, there does not appear to be much difference in the performance of the scoring rules.  However, for the lowest level of imperfection, in which $\delta=0.1$, there appears to be a notable difference with the RPS outperformed by both the ignorance and Brier scores. From this graph alone, however, it is not clear whether these differences are statistically significant.  Given that each of the scores are calculated on the same sets of forecast pairs, the scoring rules can be compared pairwise with a total of three different comparisons (ignorance vs RPS, ignorance vs Brier and RPS vs Brier).  These differences are shown as a function of $\log_{2}(n)$ in figure~\ref{figure:Diff_wins_error_bars}, with each panel representing a different value of $\delta$. The error bars represent $95$ percent resampling intervals of the mean difference and hence, if the intervals do not contain zero, there is a significant difference in the performance of that pair of scoring rules. \par

For the two lowest levels of imperfection ($\delta=0.01$ and $\delta=0.025$), there is a clear hierarchy in terms of the efficacy of each score in identifying the perfect forecasting system.  The ignorance score tends to outperform the Brier score which tends to outperform the RPS.  This difference is most stark for larger values of $n$.  For the two larger levels of imperfection ($\delta=0.05$ and $\delta=0.1$), the difference is less clear and, in general, there is no significant difference between the Brier score and the RPS.  The ignorance score, on the other hand, still tends to perform significantly better than both other scores.  These results therefore provide clear support for the ignorance score and little support for the RPS. \par

\section{Discussion} \label{section:discussion}
The aim of this paper is to reopen the debate surrounding the use of scoring rules for evaluating the performance of probabilistic forecasts of football matches.  The reasoning presented by Constantinou and Fenton supporting the use of the RPS over other scoring rules has been shown to be oversimplistic and the conclusion questionable.  With this in mind, two experiments have been conducted with the aim of assessing the performance of each scoring rule in the context of identifying a perfect forecasting system using a finite number of past forecasts and outcomes.  The ignorance score has been found to outperform both the RPS and the Brier scores whilst, to a lesser extent, the Brier score has been shown to perform better than the RPS in this context.  \par

The results in this paper may seem surprising at first.  After all, both the Brier score and the RPS are non-local and take into account the entire forecast distribution rather than just the probability at the outcome whilst the RPS is sensitive to distance and therefore also takes into account the ordering of the potential outcomes.  It would be easy to conclude from this that, since both scores take more of the distribution into account, they are more informative.  However, it should be stressed that this would only be the case if those extra aspects are genuinely useful in terms of assessing the performance of the forecasts.  In practice, we only ever gain limited knowledge regarding the true distribution, even once the outcome is revealed.  If, for example, the outcome is a home win, this tells us little or nothing about the probability of a draw or an away win.  In fact, knowing the outcome reveals relatively little about its probability, other than it is greater than zero.  Given this, we argue that the probability placed on potential outcomes that didn't happen are irrelevant. We know nothing about the true probabilities and therefore cannot reward probability placed on such outcomes.  On the other hand, we \emph{know} that the actual outcome occurred.  Moreover, the more likely that event was deemed by the forecast, the better prepared we could have been for the occurrence of that outcome.  We therefore argue that the probability placed on the outcome can be the only aspect of interest in evaluating probabilistic forecasts.  Given that the ignorance score is the only proper and local score (\cite{brocker2007scoring}), this leads to it being a natural preference. \par 

In summary, this paper has both argued for and provided empirical evidence in favour of the ignorance score over both the Brier score and RPS.  It should be noted, however, that this paper has only touched upon the question of which types of imperfect forecasts are favoured by different scores.  Useful future work would be to attempt to understand better where different scoring rules favour different types of forecasts.  A preference for the types of forecasts favoured by the Brier score or RPS would then need to be weighed up against the unfavourable results demonstrated in this paper.  Regardless, we hope that the arguments and results in this paper are successful in reopening the debate surrounding the choice of scoring rule for evaluating forecasts of football matches.  From the evidence presented in this paper, we strongly recommend the ignorance score for this purpose. \par

\newpage
\appendix

\section{Data}
Experiment two makes use of bookmakers' odds on actual football matches to form probabilistic forecasts.  These odds are taken from the data set available at \url{www.football-data.co.uk} which supplies free-to-access match-by-match data on 22 European Leagues dating back as far back as the 1993/1994 season. Here, data from the top five English leagues are used and are summarised in table~\ref{table:Leagues_available}.   \par

\begin{table}[!htb]
\begin{center}
\begin{tabular}{|l|c|c|}
\hline
League & First available season & Number of matches \\ 
\hline
English Premier League & 2005/2006 & 6460 \\ 
English Championship & 2005/2006 & 6624 \\
English League One & 2005/2006 & 6624 \\  
English League Two & 2005/2006 & 6624 \\
English National League & 2005/2006 & 6488 \\  
\hline
\end{tabular}
\caption{Football league data used in this paper.}
\label{table:Leagues_available}
\end{center}
\end{table}

\section{Match probabilities from odds}
Let $O_{h}$, $O_{d}$ and $O_{a}$ be the decimal odds on a home win, draw and away win respectively for a given match. The multiplicative inverse of the odds on each outcome represents the `implied' probability.  However, due to the profit margin of the bookmakers, the implied probabilities will generally sum to a value greater than one.  To remove the profit margin, the implied probabilities are divided through by their sum and therefore a probabilistic forecast is formed by 
\begin{equation} \label{eq1}
\begin{split}
p_{h} & = \frac{1/O_{h}}{1/O_{h}+1/O_{d}+1/O_{a}}, \\
p_{d} & = \frac{1/O_{d}}{1/O_{h}+1/O_{d}+1/O_{a}}, \\
p_{a} & = \frac{1/O_{a}}{1/O_{h}+1/O_{d}+1/O_{a}}. \\
\end{split}
\end{equation}
Forecasts are formed using the maximum odds (over all bookmakers given) in each case.

\newpage
\bibliographystyle{agu}
\bibliography{bibliography}
\end{document}